\documentclass[%
 reprint,
 amsmath,amssymb,
 aps,
]{revtex4-1}

\usepackage{graphicx}
\usepackage{dcolumn}
\usepackage{bm}

\begin{document}

\preprint{APS/123-QED}

\title{Epidemic spreading induced by diversity of agents' mobility}

\author{Jie Zhou}
 \affiliation{Temasek Laboratories, National University of
Singapore, Singapore 117411.}
\author{Ning Ning Chung}%
\affiliation{%
 Temasek Laboratories, National University of
Singapore, Singapore 117411.
}%

\author{Lock Yue Chew}
\affiliation{
 Division of Physics and Applied Physics, School
of Physical and Mathematical Sciences, Nanyang Technological
University, 21 Nanyang Links, Singapore 637371.
}%

\author{Choy Heng Lai}
\affiliation{Temasek Laboratories, National University of
Singapore, Singapore 117411.}

\affiliation{Beijing-Hong Kong-Singapore Joint Centre for
Nonlinear and Complex Systems (Singapore), National University of
Singapore, Kent Ridge, Singapore 119260.}

\affiliation{Department of Physics, National University of
Singapore, Singapore 117542.}


\begin{abstract}
In this paper, we study into the impact of the preference of
an individual for public transport on the spread of infectious
disease, through a quantity known as the public mobility. Our
theoretical and numerical results based on a constructed model
reveal that if the average public mobility of the agents is
fixed, an increase in the diversity of the agents' public
mobility reduces the epidemic threshold, beyond which an
enhancement in the rate of infection is observed. Our findings
provide an approach to improve the resistance of a
society against infectious disease, while preserving the
utilization rate of the public transportation system.
\begin{description}
\item[PACS numbers] 89.75.Hc, 05.45.Xt, 89.75.Fb.
\end{description}
\end{abstract}

\pacs{Valid PACS appear here}
\maketitle


\section{Introduction}
\label{sec:Introduction}

The advent of modern transportation systems has enhanced the
mobility of mankind and has increased their range of travel. At
the same time, it has intensified the contact between human beings
because of the higher human density within transportation systems
resulting from a confluence of people within limited physical
spaces. The close proximity between travellers provides an
opportunity for diseases to spread and it is well known that
infectious disease is the main cause of death, disability, as well
as social and economic disruption that affects millions of people
\cite{www1,Barquet:1997,Anderson:1991,Bailey:1993}. To stop the
proliferation of infectious disease and their spread, researchers have searched for
ways to hinder their diffusion \cite{Dezso:2002,Chen:2008,Shaw:2010,www2,Salathe:2010}. A typical strategy is to adjust the level of human contact through the temporary closure of companies and educational institutes. This strategies, however comes at a very high price for both society and economy.

In this paper, we have focused our research on epidemic spreading
in public transportation system, with the aim of understanding how
the disease spreads within such a system so that mitigating strategies
can be  determined to reduce its social and economic impact.
Human mobility relates to the activity of moving from one point in
space to another and can be measured by the frequency and distance
of travel \cite{Brockmann:2005,Gonzalez:2008,Balcan:2009,Song:2010,Neal:2011}.
Such human movements have been enhanced by public transportation
network which is an indispensable component of the major metropolitan
area of a country. For example, the Mass Rapid Transit (MRT) system
in Singapore has a daily load of around $700,000$ passengers
(i.e. $15\%$ of the total population) \cite{Xiuju:2009}. About $90\%$
of Hong Kong citizens rely on public transport facilities for commuting
with the main concern being exposure to airborne pollutants within the public vehicles \cite{Chan:2002}. The large flux of commuters
in public transportation system has typically led to extreme
overcrowding, especially during peak hours. The resulting high
rate of human contact implies a high rate of transmissibility of
infectious diseases. For example, the risk of contracting pulmonary
tuberculosis in Peru is higher by a factor of $4.09$ for those
commuting by minibus compared to those traveling by private
transportation \cite{Horna:2007}. In consequence, commuters tend to
avoid public transportation during an epidemic outbreak. They choose
either to stay at home or to commute by private transport. The outcome
is undesirable: a severe shortage of manpower in the workplace and the
possible occurrence of major traffic congestion.

In the past decades, there is a lot of interest in studying the
spread of epidemics within complex networks, which includes: (i) the
influence of network structure on epidemic spreading \cite{Kuperman:2001,Vespignani:2001,Vespignani:2003,Barthlemy:2004,Yan:2005},
(ii) the development of immunization
strategies \cite{Dezso:2002,Zanette:2002,Cohen:2003,Gallos:2007,Chen:2008,Shaw:2010},
and (iii) epidemic spreading in community networks \cite{Liu:2005,Zhou:2007,Zhou:2009},
in dynamic networks \cite{Frasca:2006,Fefferman:2007,Volz:2007,Buscarino:2008} and in adaptive
networks \cite{Gross:2006&2008,Shaw:2008,Zanette:2008,Zhou:2012}.
This has motivated us to construct a model on epidemic spreading that relates to public transportation system. It is known that commuters have diverse preference in choosing their mode of travel
\cite{Davidov:2003} based on their socioeconomic status. As a result, the frequency of different agents
using the public transportation system may differ. In this paper,
we shall denote the frequency of utilizing public transport for travel
as ``public mobility". Individuals with high public mobility
use public transports very often, while individuals with low
public mobility hardly use the public transportation system. This
dichotomy in the usage of public transport prompts the following
question\,: how does the diversity of public mobilities affect the
speed of epidemic spreading? The purpose of this work is to give a
definite answer to this question.

The structure of our paper is as follow. The details of the model is discussed in
Sec.\,\ref{sec:Model} of this paper. In Sec.\,\ref{sec:Theoretical analysis}
of the paper, we provide a theoretical analysis that enables us to determine
the lower bound of the epidemic threshold. Then, in
Sec.\,\ref{sec:Simulation results}, we present simulation results which
are found to support our theoretical analysis. Finally, we end
our paper with a discussion and conclusion in
Sec.\,\ref{sec:Discussion and conclusion}.

\section{Model}
\label{sec:Model}

In order to gain a better understanding on our model, it is useful
to first study a simple model, which can be regarded as a null model
that serves the purpose of a benchmark and validity check. In this
simple model, a square with length $L$
satisfying periodic boundary condition is
used to represent a society. There are $N$ agents in the square.
The positions of the agents are randomly assigned with a uniform
distribution and for the sake of simplicity, are assumed to be
fixed over time. We also assume that agent only interacts with agents
who are located at a distance that is less than $r$ away. In other
words, links appear between all pairs of agents whose distance
from each other is smaller than $r$. Since the agents are fixed in
their position, the links that are established in this way do
not change with time. This null model is simple, and its similar forms have also been adopted in other works \cite{Zhou:2009,Frasca:2006}.
In this model, the average degree of the network of agents in the
square: $\langle k\rangle$, which is defined as the average number
of links that an agent has, is approximately given by $N\pi r^2/L^2$,
and the degree distribution of this model satisfies
the binomial distribution $p(k)=C_N^kq_L^k(1-q_L)^{N-k}$ with $q_L=N\pi r^2/L^2$. As the ratio
of the standard deviation of the distribution to its mean value
tends to zero when $N\rightarrow\infty$, we expect the degree of
the network connection between agents to be homogenous.

\begin{figure}[b]
\includegraphics[width=1.0\linewidth]{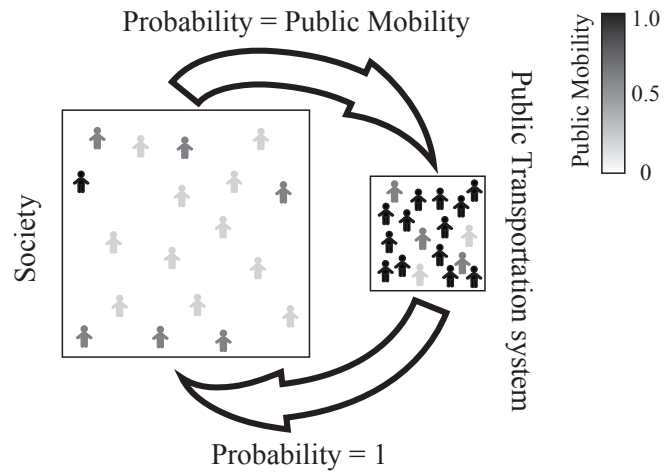}
\caption{\label{Fig:Illustration} Schematic illustration of the
model. The smaller square represents the public transportation
system and the larger square represents the rest of the society.
At the beginning of each time step, each agent may transit to the
public transportation system with a probability that equals his
public mobility. At the end of each time step, all the agents in
the public transportation system return to the society. Note that
an agent in darker gray possesses a higher public mobility.}
\end{figure}

In this paper, we use the SIS model to describe the
epidemiological process, which is widely adopted to describe
infectious diseases
\cite{Weiss:1971,Bailey:1975,Murray:1993,Diekmann:2000}. In this
model, agents can be in either of two distinct states: susceptible
or infected. A susceptible agent may become infected if there are
infected agents within the interaction radius. Suppose a
susceptible agent has $k$ neighbors within its interaction region,
of which $k_\mathrm{inf}$ are infected, and the probability of
being infected by each infected neighbor is $p$, then the
probability that agent becomes infected is $[1 - (1 -
p)^{k_\mathrm{inf}}]$. At the same time, each infected agent can
recover from the disease and becomes susceptible. We assume that
this occurs at a rate of $\mu$. When the ratio $p/\mu$ is fixed,
different pairs of $p$ and $\mu$ only affect the definition of the
time scale of the disease propagation \cite{Diekmann:2000}.
Therefore, we can set $p$ and $\mu$ to be sufficiently small so
as to use the approximation
$[1 - (1 - p)^{k_\mathrm{inf}}]\rightarrow p\,k_\mathrm{inf}$.
In other words, we can maintain the same results (except
for the time scale) as long as the ratio $p/\mu$ is
fixed. This approximation has been widely adopted in the
literatures (see Ref.\,\cite{Zanette:2008,Zhou:2006}).

In this paper, we have fixed $r = 0.02$, $p = 0.1$, $\mu = 0.2$.
In the simulations, all the averaged results and
their standard deviation (which is indicated by the error-bars), are obtained from $1000$ different realizations, if not otherwise specified.

The epidemic threshold of the null model is determined by the basic reproductive number $R_0$ with \cite{Anderson:1991,Fraser:2009}
\begin{equation}
\label{eq:R0} R_0=p\langle{k}\rangle/\mu\,.
\end{equation}
When $R_0<1$ the infection dies out in the long run, and when $R_0>1$ the infection may spread over the population. This condition leads to a critical average degree $\langle{k}\rangle^\mathrm{th}=\frac{\mu}{p}$ and correspondingly the critical number $N^\mathrm{th}$ for a given set of $r$, $p$, $\mu$ and $L$ with $N^\mathrm{th}=\frac{\mu}{p}\frac{L^2}{\pi r^2}$.

Now we are ready to introduce our model which focuses primarily on
the public transportation system. In order to study epidemic spreading in
public transportation system, we have separated the society into two
parts: the public transportation system ($A$) and the rest of the
society ($B$) (see Fig.\,\ref{Fig:Illustration}). Each part
is represented by a square which satisfies periodic boundary
condition. The length of square $A$ is $L_A$ and the length of
square $B$ is $L_B$. Since human contacts within public
transportation system is typically denser, we have set $L_A \ll
L_B$. There are a total of $N$ agents in the society, each
(labelled by the index $e$) with a public mobility (denoted as
``PM" in the following) of $m_e$ that ranges
between $0$ and $1$. Our model begins by assigning a random
position for each of the $N$ agents in square $B$. At the
beginning of each time step, agent $e$ either transits to square
$A$ with probability $m_e$ and then chooses a random
position there to stay, or remains in square $B$ at the originally
assigned position. At the end of each time step, all the agents
that transit to square $A$ return to their original position in
square $B$ and the whole system prepares for the next time
step.

Similar to the null model, each agent has a contact radius of $r$
and a link between two agents is formed whenever they are within
this radius. The average degree of agents in square $A$ ($B$) is
$\langle k\rangle_A=N_A\pi r^2/L_A^2$ ($\langle k\rangle_B=N_B\pi
r^2/L_B^2$), where $N_A$ ($N_B$) is the number of agents in the
square $A$ ($B$).

\section{Theoretical analysis}
\label{sec:Theoretical analysis}

In this section, we provide a theoretical analysis on the epidemic
threshold of our model. Suppose $\rho(m)$ is the fraction of
agents with PM $m$ such that $\int\rho(m)dm=1$. Then, the average
PM $\overline{m}=\int \rho(m)mdm$ and the second moment $D=\int
\rho(m)m^2dm$. Let us denote $i(m,t)$ as the fraction of infected
nodes with PM $m$ at time step $t$. Since $i(m,t)$ represents the
fraction of infected agents in both square $A$ and $B$ at time
$t$, we expect the evolution of ${i(m,t)}$ to consist of two parts:
\begin{equation}
\label{eq:2areas} i(m,t+1)=\Pi_A(m,t+1)+\Pi_B(m,t+1)\,.
\end{equation}
Here, $\Pi_A(m,t+1)$ denotes the fraction
of infected agents with PM $m$ at time $t+1$ as a result of having
visited square $A$ at time $t$. It can be expressed as follow:
\begin{eqnarray}
\label{eq:PiA} \Pi_A(m,t+1)&=&i(m,t)\cdot
m\nonumber\\&&+[\rho(m)-i(m,t) ]\cdot m\cdot p
\bar{k}_\mathrm{inf}^{(A)}(t)\nonumber\\&&-\mu i(m,t)\cdot m \,,
\end{eqnarray}
where $\bar{k}_\mathrm{inf}^{(A)}(t)$ is the average number of
infected neighbors of an agent in square $A$ at time $t$ and
is defined by:
\begin{equation}
\label{eq:DegA} \bar{k}_\mathrm{inf}^{(A)}(t)=\langle
k\rangle_A\frac{\int i(m',t)m'\mathrm{d}m'}{\int \rho(m')m'\mathrm{d}m'} \,.
\end{equation}
The first term on the right hand side of Eq.\,(\ref{eq:PiA})
denotes the fraction of infected agents with PM $m$ that have
moved to square $A$ at time $t$. The second term on the right hand
side denotes the fraction of susceptible agents with PM $m$ who
are infected at time $t$ due to their transit to square $A$. The
third term on the right hand side represents the fraction of
infected agents with PM $m$ who have recovered from the infected
state and are now in the susceptible state, as they traverse to
square $A$ at time step $t$. On the other hand, $\Pi_B(m,t+1)$
gives the fraction of infected agents with PM $m$ at time $t+1$ in
lieu of having remain in square $B$ at time $t$. Similar to
$\Pi_A(m,t+1)$, $\Pi_B(m,t+1)$ takes the following form:
\begin{eqnarray}
\label{eq:PiB} \Pi_B(m,t+1)&=&i(m,t)\cdot
(1-m)\nonumber\\&&+[\rho(m)-i(m,t)]\cdot (1-m)\cdot p
\bar{k}_\mathrm{inf}^{(B)}(t)\nonumber\\&&-\mu i(m,t)\cdot(1-m)\,,
\end{eqnarray}
where $\bar{k}_\mathrm{inf}^{(B)}(t)$ is the average number of
infected neighbors in contact with an agent in square $B$ at time
$t$ and it is defined by:
\begin{equation}
\bar{k}_\mathrm{inf}^{(B)}(t)=\langle k\rangle_B\frac{\int
i(m',t)(1-m')\mathrm{d}m'}{1-\int \rho(m')m'\mathrm{d}m'} \,.
\end{equation}

In the steady state, we expect $i(m,t+1)=i(m,t)=i^*(m)$. Thus, we
have
\begin{eqnarray}
\label{eq:stable} \frac{\mu}{p} i^*(m)&=&[\rho(m)-i^*(m)]\cdot m\cdot\bar{k}_\mathrm{inf}^{(A)}(t)\nonumber\\
&&+[\rho(m)-i^*(m)]\cdot\bar{k}_\mathrm{inf}^{(B)}(t)\nonumber\\
&&-[\rho(m)-i^*(m)]\cdot m\cdot\bar{k}_\mathrm{inf}^{(B)}(t)\,.
\end{eqnarray}
Multiplying both sides of the equation by $m$ and then integrating
throughout with respect to $m$, we obtain
\begin{eqnarray}
\label{eq:T3} & &\Omega\left[\frac{\langle
k\rangle_A}{\overline{m}}+\frac{\langle
k\rangle_B}{1-\overline{m}}(\lambda-1)(\eta-1)\right]\nonumber\\
&=&D\left[\frac{\langle k\rangle_A}{\overline{m}}-\frac{\langle
k\rangle_B}{1-\overline{m}}(\lambda-1)\right]+\frac{\langle
k\rangle_B}{1-\overline{m}}\overline{m}(\lambda-1)-\frac{\mu}{p}
\,,\nonumber\\
\end{eqnarray}

where $\lambda=i^*/\Theta$ and $\eta=\Theta/\Omega$ for $i^*=\int i^*(m)\mathrm{d}m$, $\Theta=\int i^*(m)m\mathrm{d}m$ and $\Omega=\int i^*(m)m^2\mathrm{d}m$. It is
easy to see that $\lambda-1\geq0$ and $\eta-1\geq0$ when $m\in [\,0, 1]$. Thus, the
terms in the square bracket on the left hand side of
Eq.\,(\ref{eq:T3}) is positive. In order for the fraction of
infected agents to be non-zero, i.e. $i^*>0$ and $\Omega>0$, the
right hand side of Eq.\,(\ref{eq:T3}) has to be positive. Hence,
\begin{eqnarray}
\label{eq:T4} D\left[\frac{\langle
k\rangle_A}{\overline{m}}-\frac{\langle
k\rangle_B}{1-\overline{m}}(\lambda-1)\right]+\frac{\langle
k\rangle_B}{1-\overline{m}}\overline{m}(\lambda-1)>\frac{\mu}{p}\,.\nonumber\\
\end{eqnarray}
Eq.\,(\ref{eq:T4}) indicates the presence of a lower bound for
$D$, which is
\begin{eqnarray}
\label{eq:DLambda}
\underline{D}=\dfrac{\dfrac{\mu}{p}-\dfrac{\langle
k\rangle_B}{1-\overline{m}}\overline{m}(\lambda-1)}{\dfrac{\langle
k\rangle_A}{\overline{m}}-\dfrac{\langle
k\rangle_B}{1-\overline{m}}(\lambda-1)}\,.
\end{eqnarray}

When $\langle k\rangle_B\ll\langle k\rangle_A$ and $\langle k\rangle_B$
is small, Eq.\,(\ref{eq:DLambda}) can be approximated by
\begin{equation}
\label{eq:Dka}
\underline{D}=\dfrac{\mu}{p}\dfrac{\overline{m}}{\langle
k\rangle_A}\,.
\end{equation}
Since $\langle k\rangle_A=\frac{N_A\pi
r^2}{L_A^2}=\frac{\overline{m}N\pi r^2}{L_A^2}$, $\underline{D}$
can also be expressed in terms of $N$ and $L_A$. This expression has the implication that the value of $\underline{D}$
remains unchange as we scale the variables $N$ and $L_A^2$ by the
same factor.

Let us next consider the case where $\langle k\rangle_B$ is not
neglected in Eq.\,(\ref{eq:DLambda}). Since human contacts in
public transportation system is typically denser, we anticipate square
$A$ to dominate the infection process. Therefore, agents that transit
frequently to square $A$ have a larger probability of being infected.
Hence, we expect $i^*(m)\sim \rho(m)m$, and therefore
$\lambda={i^*}/{\Theta}={\int\rho(m)mdm}/{\int \rho(m)m^2dm}
={\overline{m}}/{D}$. Thus, Eq.\,(\ref{eq:DLambda}) becomes
\begin{eqnarray}
\label{eq:DD} \underline{D}=\dfrac{\dfrac{\mu}{p}-\dfrac{\langle
k\rangle_B}{1-\overline{m}}\overline{m}\left(\dfrac{\overline{m}}
{\underline{D}}-1\right)}{\dfrac{\langle
k\rangle_A}{\overline{m}}-\dfrac{\langle
k\rangle_B}{1-\overline{m}}
\left(\dfrac{\overline{m}}{\underline{D}}-1\right)} \,.
\end{eqnarray}
Solving Eq.\,(\ref{eq:DD}) for $\underline{D}$\,, we obtain

\begin{eqnarray}
\label{eq:DSolution}
\underline{D}&=&\dfrac{\sqrt{\left(\dfrac{\mu}{p}+2\overline{m}\dfrac{\langle
k\rangle_B}{1-\overline{m}}\right)^2
-4\overline{m}^2\dfrac{\langle
k\rangle_B}{1-\overline{m}}\left(\dfrac{\langle
k\rangle_A}{\overline{m}}+\dfrac{\langle
k\rangle_B}{1-\overline{m}}\right)}}{2\left(\dfrac{\langle
k\rangle_A}{\overline{m}}+\dfrac{\langle
k\rangle_B}{1-\overline{m}}\right)}\,\nonumber\\
&&+\dfrac{\left(\dfrac{\mu}{p}+2\overline{m}\dfrac{\langle
k\rangle_B}{1-\overline{m}}\right)}{2\left(\dfrac{\langle
k\rangle_A}{\overline{m}}+\dfrac{\langle
k\rangle_B}{1-\overline{m}}\right)}\,.
\end{eqnarray}
By using $\langle k\rangle_A=\frac{\overline{m}N\pi r^2}{L_A^2}$
and $\langle k\rangle_B=\frac{(1-\overline{m})N\pi r^2}{L_B^2}$,
we can also express $\underline{D}$ via $N$, $L_A$ and $L_B$.

Similar to Eq.\,(\ref{eq:Dka}), $\underline{D}$ is found here
to remain invariant when $N$, $L_A^2$ and $L_B^2$ are varied by the same
scaling factor if $\overline{m}$ is fixed. In fact, Eqs.\,(\ref{eq:Dka}) and (\ref{eq:DSolution})
allow us to obtain the lower bound of the variance of PM:
\begin{equation}
\label{eq:sigmaLB}
\underline{\sigma^2}=\underline{D}-\overline{m}^2\,.
\end{equation}

\section{Simulation results}
\label{sec:Simulation results}

Before we show the influence of the diversity of PM on epidemic
spreading, let us first study the simple case when all
the agents have the same PM value $m$. We suppose the system
contains $N$ agents. When the PM of all the agents is $m$, the number
of agents in square $A$ in each time step is $N_A=mN$, while that in
square $B$ is $N_B=(1-m)N$. Therefore, the long time average of the
degree of all the agents is $\langle k\rangle=m \cdot
\frac{mN\pi r^2}{L_A^2}+(1-m)\cdot\frac{(1-m)N\pi r^2}{L_B^2}$,
which is approximately $m^2\frac{N\pi r^2}{L_A^2}$ when $L_A \ll
L_B$. By adopting Eq.\,(\ref{eq:R0}), we can obtain the threshold
$m^\mathrm{th}$ as follow:
\begin{equation}
\label{eq:mth} m^\mathrm{th}=\sqrt{\frac{\mu}{p}\frac{L_A^2}{N\pi
r^2}}\,.
\end{equation}
Note that above this threshold, the system may become endemic.

\begin{figure}[b]
\includegraphics[width=1.0\linewidth]{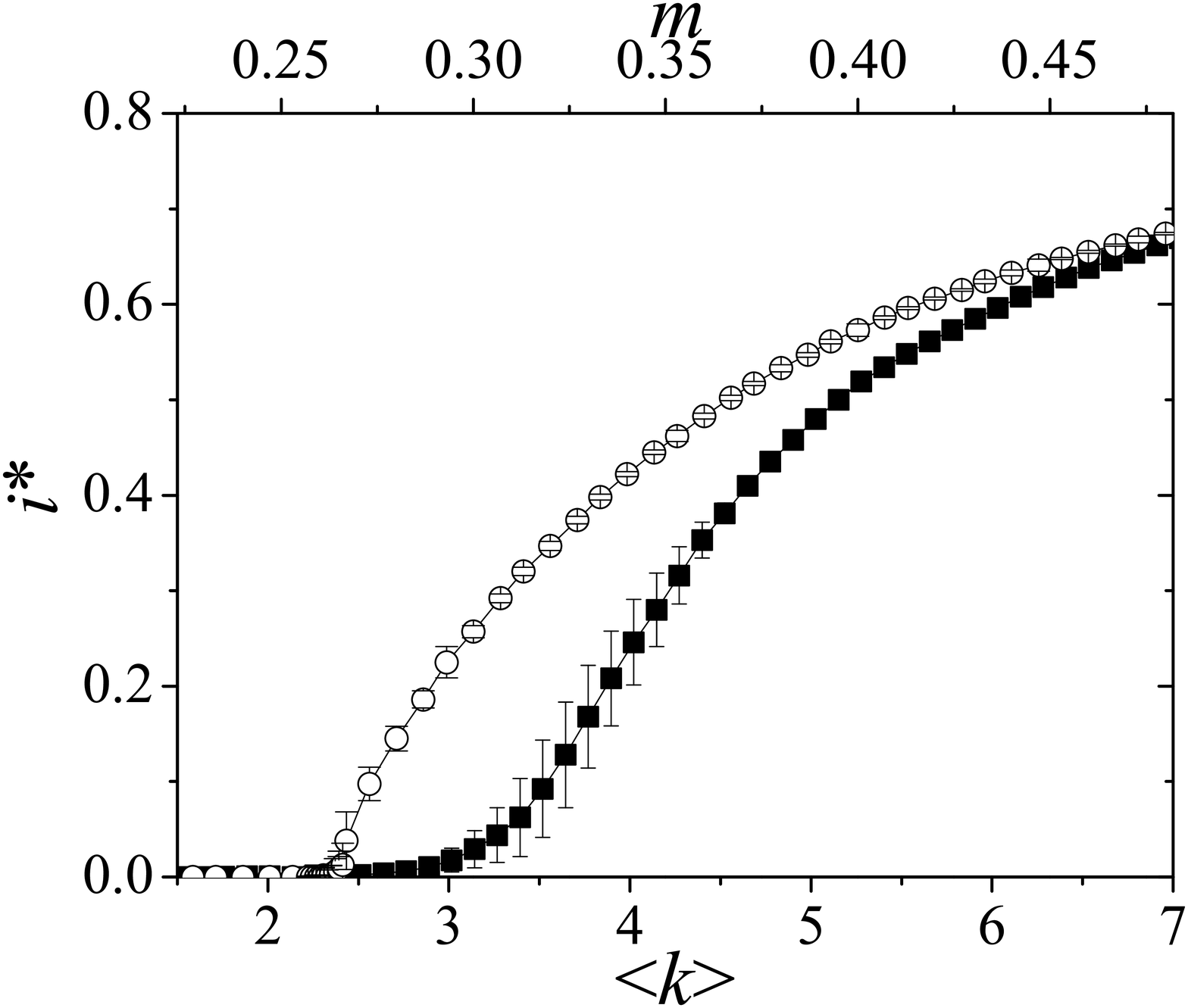}
\caption{\label{Fig:m1=m2} The open circle symbols shows the fraction of infected agents $i^*$
at the steady state versus the average degree $\langle k\rangle$ (the lower abscissa) and the average PM $m$ (the upper abscissa) for the case that the PM of all the agents are equal to $m$, with $N=1500$, $L_A=0.25$ and $L_B=10$. The solid square symbols illustrate the case for the null model for the sake of comparison, with $L=1$.}
\end{figure}

Figure \ref{Fig:m1=m2} shows the fraction of infected agents $i^*$
at the steady state versus the value $m$ and $\langle k\rangle$ for the case when the PM
of all the agents is $m$ (circle symbols) and the null model (square symbols), respectively. By using Eqs.\,(\ref{eq:R0}) and (\ref{eq:mth}), we have
$\langle k\rangle^\mathrm{th}=2$ and $m^\mathrm{th}=0.26$ (detailed parameters are indicated in the caption of the figure). It shows that our theoretical estimates on $\langle k\rangle^\mathrm{th}$ and $m^\mathrm{th}$ are in accord with the simulation results. Moreover, it shows that when $\langle k\rangle>\langle k\rangle^\mathrm{th}$, the value of the circle symbols may be much larger than that of the square symbols, which indicates that the extent of the epidemic prevalence is strongly enhanced by the public transportation system. In the following, we shall show
that a modification to this threshold behavior can occur
when we take the diversity of the agents' PM into consideration.

Let us begin by considering the case in which all the agents belong to
either of two groups: $G_1$ or $G_2$, with all the agents in each
group having the same PM. The PM of the agents in $G_1$ ($G_2$) is
$m_1$ ($m_2$), and the size of the group is $N_1$ ($N_2$). Thus,
the size of the system $N=N_1+N_2$, the average PM
$\overline{m}=(N_1\cdot m_1+N_2\cdot m_2)/N$, and the variance of
the PM $\sigma^2=(N_1\cdot m_1^2 +N_2\cdot
m_2^2)/N-\overline{m}^2$. At each time step, the expected number
of agents in $G_1$ ($G_2$) transiting to square $A$ is equal to
$m_1N_1$($m_2N_2$). Hence, on average, square $A$ contains
$N_A=m_1N_1+m_2N_2=\overline{m}N$ agents, and square $B$ contains
$N_B=(1-\overline{m})N$ agents. Therefore, $\langle k\rangle_A$
and $\langle k\rangle_B$ can be expressed as $\langle
k\rangle_A=\overline{m}N\pi r^2/L_A^2$ and $\langle
k\rangle_B=(1-\overline{m})N\pi r^2/L_B^2$, respectively.
Moreover, after the process of time averaging, the degree of agent $e$
with PM $m_e$ is $k_e=m_e\langle k\rangle_A+(1-m_e)\langle
k\rangle_B$.

\begin{figure}[b]
\includegraphics[width=1.0\linewidth]{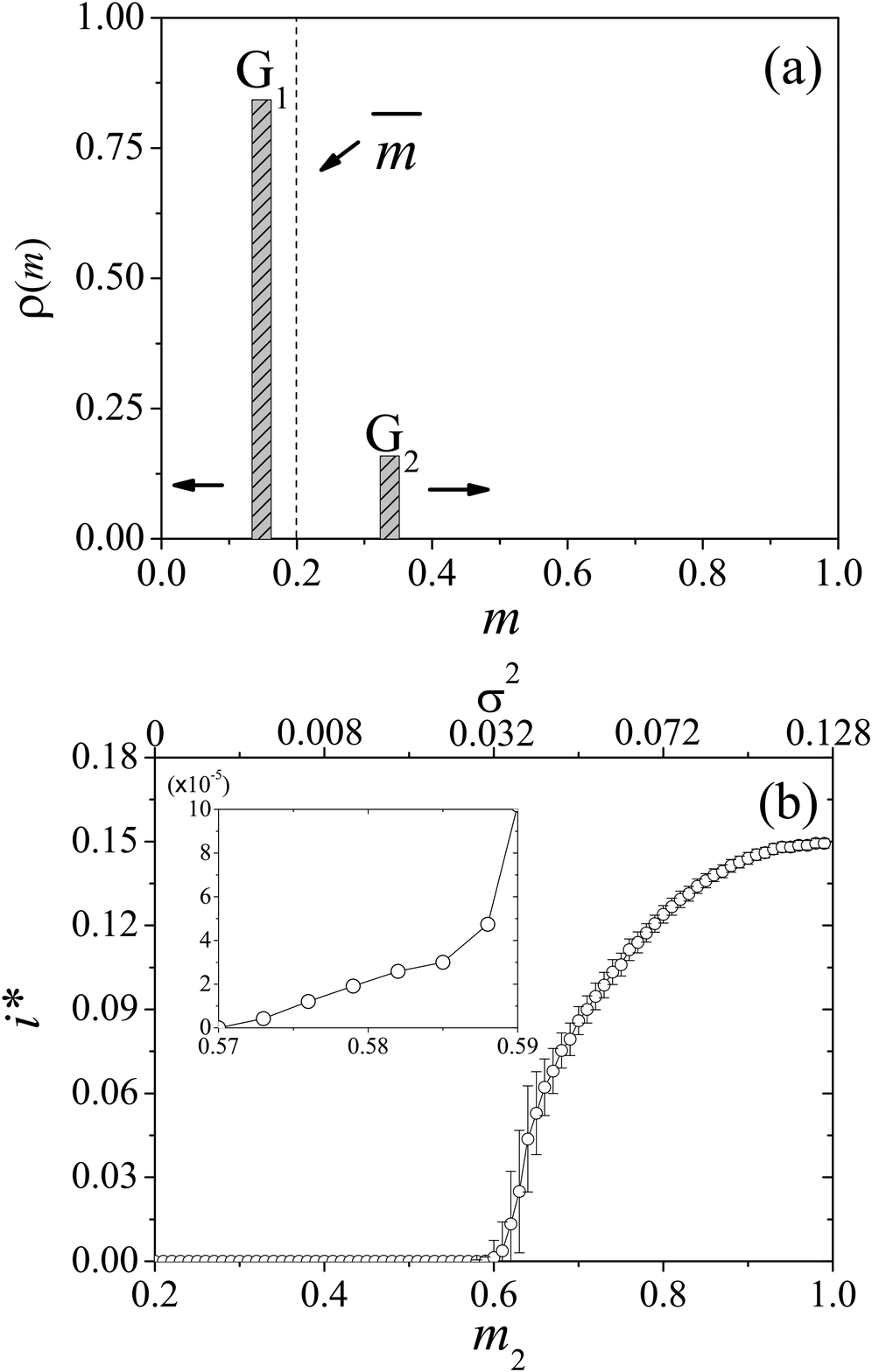}
\caption{\label{Fig:2motilities} (a) The fraction of agents $\rho$
against the PM $m$. The two bars in the figure indicate the
presence of only two groups, with each group having different PM.
This figure illustrates how the standard deviation of the PM is
tuned without changing the average PM, which is
$\overline{m}=0.2$, as indicated by the dashed line. (b) The
fraction of infected agents $i^*$ at the steady state against the
PM $m_2$ and $\sigma^2$. The inset shows the detailed behavior of $i^*$ near the zone of transition. In this case, $N_1=1250$ and
$N_2=250$. Note that all the other parameters take the same value as those employed in Fig.\,\ref{Fig:m1=m2}.}
\end{figure}

In order to demonstrate the effect of diversity in PM on epidemic
spreading, we first set $m_1=m_2$. After that, we decrease $m_1$
and increase $m_2$ such that $\overline{m}$ remains unchanged.
This operation increases $\sigma^2$ from $0$ without changing
$\overline{m}$. Fig.\,\ref{Fig:2motilities}(a) illustrates the
manner in which the PM of the two groups are tuned.
Fig.\,\ref{Fig:2motilities}(b) shows the fraction of infected
agents $i^*$ at the steady state as a function of $m_2$. When $m_1=m_2=0.2$ (i.e.,
$\sigma^2=0$), the system is in a disease free state. When $m_2$
exceeds the threshold: $m_2^\mathrm{th}\sim0.568$ which is determined from Eq.\,({\ref{eq:Dka}}) (note that the rest of the parameters are indicated in the caption of the figure), the system
becomes endemic. Compared with the results shown in
Fig.\,\ref{Fig:m1=m2} and Eq.\,({\ref{eq:mth}}), this example
shows that the diversity of the PM can induce epidemic
spreading, even when $\overline{m}$ is smaller than the
$m^\mathrm{th}$ in Eq.\,({\ref{eq:mth}}). Since in this
case $L_A\ll L_B$ which makes $\langle k \rangle_A\gg\langle k
\rangle_B$, we can use Eq.\,(\ref{eq:Dka}) to calculate the lower
bound of the variance $\underline{\sigma^2}$. Our calculation
gives $\underline{D}=0.067$, and correspondingly
$\underline{\sigma^2}=\underline{D}-\overline{m}^2=0.027$.
This figure shows that our simulation results are in accord with the theoretical estimate on the epidemic threshold, above which a finite fraction
of the infected agents is found to exist. Thus, for a fixed
$\overline{m}$, there exists a threshold for the variance of the
PM, exceeding which the epidemic spreads and the system becomes
endemic. Beyond the threshold, the fraction of infected agents
increases as $\sigma^2$ increases.

In a more general setting, the PM of the population may follow an
arbitrary distribution. In order to study our model
in this more general situation, we first need to develop an
approach to assign PM to agents following a given distribution,
with $\overline{m}$ and $\sigma^2$ tunable in this method. The
details of this method are presented in the appendix.

\begin{figure}[b]
\includegraphics[width=1.0\linewidth]{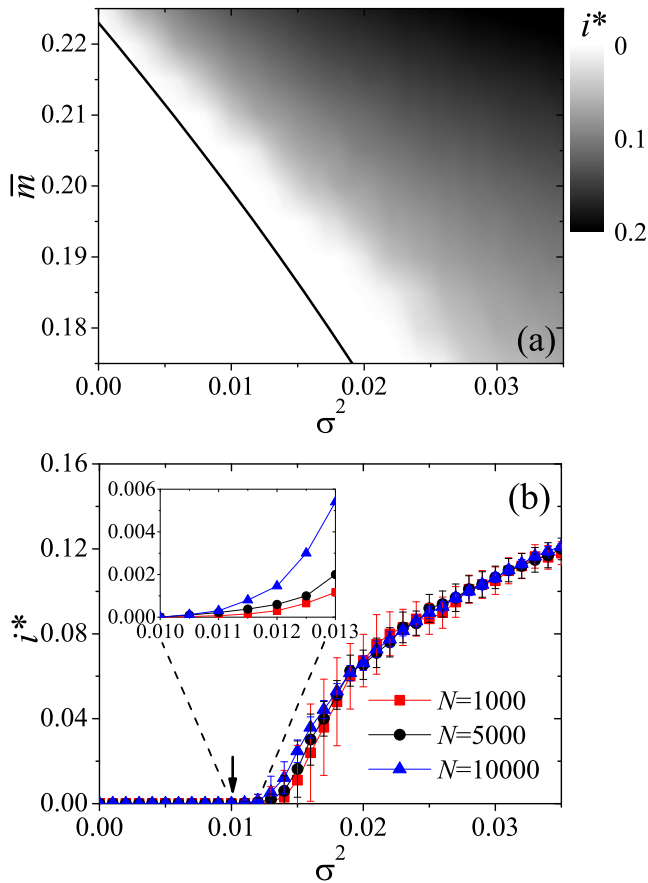}
\caption{\label{Fig:mD} (Color online) (a) A gray-scale plot on
the fraction of infected agents $i^*$ at the steady state in the
$\overline{m}$-$\sigma^2$ plane. The sizes of the error-bars are small and hence are not shown here.
The parameters employed are: $\alpha=3$, $N=2000$, $G=20$, $L_A=0.25$
and $L_B=5$. (b) The fraction of infected agents $i^*$ at the steady
state as a function of $\sigma^2$ when $\overline{m}=0.2$ for
$N=1000$, $5000$ and $10000$. In all the three cases, $N/L_A^2$ and
$N/L_B^2$ are maintained at the same value. That is (i) $N=1000$,
$L_A\simeq0.177$, $L_B\simeq3.54$, (ii) $N=5000$, $L_A\simeq0.4$,
$L_B\simeq7.9$, (iii) $N=10000$, $L_A\simeq0.56$, $L_B\simeq11.2$.
As a result, $\langle k \rangle_A\simeq8$ and
$\langle k \rangle_B\simeq0.08$ for all the three cases. The inset shows the zoom in results for the range $\sigma^2\in[0.010\,,0.013]$. The results in this panel are obtained from $5000$ different realizations.
Note that the values of the other parameters are the same as those
used in (a).}
\end{figure}

With the observation that the commuters behavior \cite{Neal:2011} and their traveling properties such as distance and time interval between journeys are found to be characterized by
power-law distribution
\cite{Brockmann:2005,Gonzalez:2008,Balcan:2009,Song:2010}, we
here assume that PM follows the distribution: $\rho(m)\sim
m^{-\alpha}$. (Note that our conclusions do not rely on any particular form of $\rho(m)$). By utilizing the method introduced in the appendix, we can tune $\overline{m}$ and $\sigma^2$ without changing the form of $\rho(m)$. The results on the fraction of
infected agents $i^*$ at the steady state in the $\overline{m}$ -
$\sigma^2$ plane are shown in Fig.\,\ref{Fig:mD} (a), where the exponent of the power-law distribution is $\alpha=3$. In this figure darker grey levels indicate a larger
fraction of infected agents. We observe that for a given
$\sigma^2$, $i^*$ increases with an increase in $\overline{m}$,
which can be understood as follow. A larger $\overline{m}$ means
the transit of a larger number of agents to square $A$. This
implies a higher average degree for all the agents since $\langle
k\rangle_A$ is greater than $\langle k\rangle_B$. The outcome is
an increase in the contacts between humans within the society.
Thus, epidemic threshold reduces and disease spreads more easily.
In particular, when the threshold for $\sigma^2$ drops to zero,
the system can still be endemic
as long as $\overline{m}$ is large enough. The presence of a
threshold here for $\sigma^2$ is similar to the two-mobility group
model that we have discussed earlier. Just like the two-mobility
group model, we observe that as $\sigma^2$ increases beyond a
threshold, the fraction of infected agents increases as the
variance increases. Note that the solid line in the figure is
obtained from Eqs.\,(\ref{eq:DSolution}) and (\ref{eq:sigmaLB}) which denotes our theoretical estimate of the epidemic threshold. We can see that the theoretical estimation conform with our numerical simulation results. It is important to note that the theoretical
analysis above indicates that $D \sim \Theta$ and through
Eq.\,(\ref{eq:DegA}), reveals that $D$ is proportional to the
probability that a link is infected. In other words, for a fixed
$\overline{m}$, a larger $\sigma^2$ (see Eq.\,(\ref{eq:sigmaLB}))
implies a larger probability that a link is infected. This
explains the observation as duly shown in
Figs.\,\ref{Fig:2motilities} and \ref{Fig:mD} that increasing
$\sigma^2$, i.e. the diversity of PM, invariably increases the
fraction of infected agents. It thus clarifies our inference that
diversity in PM has the effect of reducing the threshold of an
epidemic outbreak. Figure \ref{Fig:mD} (b) shows the fraction of
infected agents $i^*$ at the steady state as a function of
$\sigma^2$ when $\overline{m}=0.2$ for $N=1000$, $5000$ and
$10000$. In all the three cases, we have kept the value of
$N/L_A^2$ and $N/L_B^2$ constant. We observe that
$\underline{\sigma^2}\simeq0.01$ for all the three cases as indicated
by the black arrow, which is consistent with the analytical
results obtained from Eqs.\,(\ref{eq:DSolution}) and
(\ref{eq:sigmaLB}). Moreover, when
$\sigma^2>\underline{\sigma^2}$, $i^*$ is observed to have very
similar monotonically increasing behavior for the different number of
agents $N$.

\section{Discussion and conclusion}
\label{sec:Discussion and conclusion}

In summary, we have studied into the effects of diversity in
public mobility on epidemic spreading by proposing a model which
separates a society into two parts: the public transportation
system and the rest of the society. In our model, we have defined
public mobility as the probability of an agent in the society
who opts to take public transport at each time step. Our results
show that a larger diversity in public mobility gives rise to a
smaller epidemic threshold. Taking into account the inevitable
diversity in socio-economic status among individuals within a
population, we have come to the conclusion that if we are able to
control the diversity in human behavior, we would be able to
enhance the resistance of a society against the onslaught of a
pending epidemic. Our results show that this can be achieved
without reducing the average public mobility of a society by
encouraging the population to use both public and private
transport with uniformity and without biasedness. For example,
commuters with low public mobility are persuaded to take public
transport more regularly while commuters having high public
mobility are urged to travel  in private transport with greater
frequency. In this way, epidemic spreading can be slowed down
without causing any traffic congestion as well as any disturbance
to the proper functioning of the public transportation system.

\appendix*
\section{A method of tuning $\overline{m}$ and $\sigma^2$}
In this appendix we report the details of a method of tuning $\overline{m}$ and $\sigma^2$, while maintaining the PM of the population according to an arbitrary distribution given by $\rho(m)$ as follow.

First, we distribute all the $N$
agents evenly into $G$ groups so that there are $N/G$ agents in
each group. Agents in the same group have the same PM, while
agents in different groups may have different PM. Public mobility
of agents in group $j$ is $m_j$, with $j=1,\cdots, G$. The upper
and lower bounds of the PM are $m_\mathrm{min}=m_0$ and
$m_\mathrm{max}=m_G$, respectively. We define $F(m)$ to be the
primitive function of $\rho(m)$, such that $F(m)=\int_{m_0}^{m}
\rho(m')\mathrm{d}m'$. Then, we assign PM to the agents according to the
following recurrent relations:
\begin{eqnarray}
\label{eq:F}
F(m_{j})-F(m_{j-1})=\frac{1}{G}(F(m_{G})-F(m_{0})),\nonumber\\
\text{for $j=1,\cdots,G$}.
\end{eqnarray}
Given $m_0$ and $m_G$, each $m_j$ can be obtained by solving
Eq.\,(\ref{eq:F}) from $j=1$ to $j=G$, from which $\overline{m}$
and $\sigma^2$ can be determined. Since $\rho(m)$ is positive for
$m\in[0,1]$, $m_j$ can only increase monotonically with $j$. The
meaning of Eq.\,(\ref{eq:F}) can be understood in the following way.
Based on our definition, $\rho(m)=\left[F\left(m_j\right)-F
\left(m_{j-1}\right)\right]/\Delta m$ with $\Delta m=m_j - m_{j-1}
\rightarrow 0$. This implies that $\rho(m)=C/\left(G\Delta m\right)$,
where $C$ is a constant equals to $F\left(m_G\right)-F\left(m_0\right)$,
according to our construction. Then, for a well defined $\rho(m)$, we
would expect $G\rightarrow \infty$ as $\Delta m \rightarrow 0$. In other
words, the distribution of PM generated by our approach becomes
accurate and tends towards the distribution $\rho(m)$ as
$G\rightarrow\infty$.

Furthermore, Eq.\,(\ref{eq:F}) can be easily extended to the more
general situation of each group having a different number of
agents. Suppose the size of group $j$ is $n_j$, then
Eq.\,(\ref{eq:F}) can be generalized to
\begin{eqnarray}
\label{eq:F-generalized}
F(m_{j})-F(m_{j-1})=\frac{n_j}{N}(F(m_{G})-F(m_{0})),\nonumber\\
\text{for $j=1,\cdots,G$}.
\end{eqnarray}
However, we shall restrict our investigation to the condition of same
group size as we explore into the effect of the diversity of PM.

\begin{figure}
\includegraphics[width=1.0\linewidth]{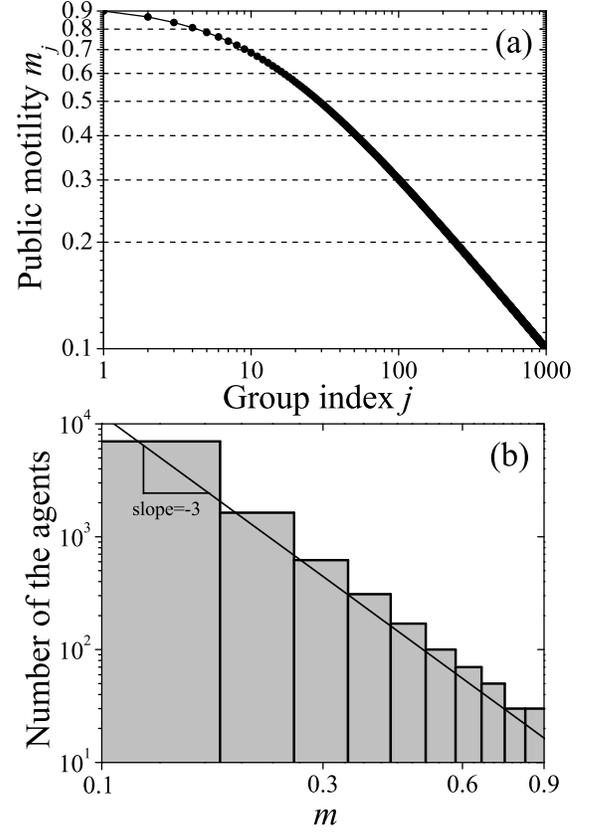}
\caption{\label{Fig:distribution} (a) The relation between the group index
$j$ and the corresponding PM $m_j$ is plotted in log-log scale,
where $\alpha=3$, $m_{0}=0.1$, $m_{G}=0.9$, $G=1000$ and
$N=10,000$. The size of each group is $N/G=10$. Dashed lines are
plotted for reference. (b) Histogram of the number of agents with
PM within the range $[m_0, m_G]$. By separating the range $[m_0,
m_G]$ into $10$ parts, the width of each bar is $0.8$. Note that
an agent with PM within the abscissa of a particular bar is
counted towards the height of that bar. The plot is in log-log
scale. We have plotted a straight line with a slope of $-3$ to
serve as a guide for reference.}
\end{figure}

Suppose PM follows the distribution: $\rho(m)\sim
m^{-\alpha}$. (Note that our formulation based on
Eqs.\,(\ref{eq:F}) and (\ref{eq:F-generalized}) allows $\rho(m)$
to take any generic form). Then, by means of Eq.\,(\ref{eq:F}), we
obtain the PM of agents in group $j$ as follow:
\begin{eqnarray}
\label{eq:power-law} m_j=\left(\frac{j}{G}\cdot
m_G^{1-\alpha}+\frac{G-j}{G}\cdot
m_0^{1-\alpha}\right)^{\frac{1}{1-\alpha}}\,.
\end{eqnarray}
The average PM is given by
\begin{eqnarray}
\label{eq:power-law-average}
&&\overline{m}=\frac{1}{G}\left(\frac{m_G^{1-\alpha}-m_0^{1-\alpha}}{G}\right)^\frac{1}{1-\alpha}\zeta\left(\frac{1}{1-\alpha},\dfrac{Gm_0^{1-\alpha}}{m_G^{1-\alpha}-m_0^{1-\alpha}},G\right)\,,\nonumber\\
&&
\end{eqnarray}
where $\zeta(\beta,p,N)=\sum_{j=1}^N(p+j)^\beta$ is the truncated
form of the generalized $\zeta$-function \cite{Newman:2005}.
Similarly, the second moment $D$ of the distribution can be obtained
from Eq.\,(\ref{eq:power-law}) as follow:
\begin{eqnarray}
\label{eq:power-law-second-moment}
&&D=\frac{1}{G}\left(\frac{m_G^{1-\alpha}-m_0^{1-\alpha}}{G}\right)^\frac{2}{1-\alpha}\zeta\left(\frac{2}{1-\alpha},\dfrac{Gm_0^{1-\alpha}}{m_G^{1-\alpha}-m_0^{1-\alpha}},G\right)\,.\nonumber\\
&&
\end{eqnarray}
We observe that $\overline{m}$ and $\sigma^2=D-\overline{m}^2$ are
functions of $m_0$, $m_G$ and $G$. Hence, we can adjust the
distribution of PM by varying the values of $m_0$ and $m_G$ so as
to obtain different $\overline{m}$ and $\sigma^2$ for a given $G$.
Figure\,\ref{Fig:distribution} (a) shows the relation of the group
index $j$ and the corresponding PM $m_j$ of a power-law
distribution with $\alpha=3$. In this case, we have set $N=10000$,
$G=1000$, $m_0=0.1$, $m_G=0.9$, $\overline{m}=0.18$ and
$\sigma^2=0.012$. In this figure, we observe that about $1\%$ of
the total number of groups has PM larger than $0.7$, while about
$90\%$ of the groups has PM smaller than $0.3$, which manifests a
strong heterogeneity in the PM. Figure\,\ref{Fig:distribution} (b)
shows the corresponding histogram of PM $\rho(m)$ of the generated
sample. We have performed a maximum likelihood estimate of the
exponent for the distribution obtained from
Eq.\,(\ref{eq:power-law}). The value of the most likely exponent
is $3.08\pm0.02$ \cite{Newman:2005}, which shows a good fit to the
value of the target exponent, which is $3$. The slight difference
between the actual and target exponent results from $G$ being
finite, as was pointed out above for a non-trivial situation, i.e.
$\alpha\neq0$, it is only when $G\rightarrow\infty$ that the
actual and target exponent have a perfect match. We have plotted a
straight line with a slope of $-3$ to serve as a guide for
reference. These results demonstrate the effectiveness of the
method. By tuning $m_0$ and $m_G$, we can obtain different values
of $\overline{m}$ and $\sigma^2$. We note that our purpose here is
not only to assign PM to the agents following a certain
distribution but more importantly to find a way to adjust
$\overline{m}$ and $\sigma^2$, hence here
$\rho(m)\,\sim\,m^{-\alpha}$ could also serve as a simple and
effective auxiliary function for adjusting $\overline{m}$ and
$\sigma^2$. In consequence, as long as $\overline{m}$ and
$\sigma^2$ are obtained correctly, we do not expect the slight
difference between the actual and target exponent to affect our
conclusions. We also note that when a power-law distribution is
unbounded, the mean and variance of the distribution can be
infinite if the exponent satisfies certain conditions
\cite{Newman:2005}. Under these circumstances, any mean and
variance obtained from a set of samples of such a distribution is
not meaningful because the fluctuation of these quantities can
become exceedingly large \cite{Newman:2005}. However, the PM in
our model is bounded within $[\,0,1]$. Therefore, the mean and
variance of the distribution obtained from
Eq.\,(\ref{eq:power-law}) are finite. In this situation, we expect
the sample mean and sample variance to converge to the population
mean and population variance of the distribution respectively, as
the number of samples tends to infinity. Thus, the mean and
variance of the distribution of our model obtained from
Eq.\,(\ref{eq:power-law}) are reliable.

\acknowledgments

This work is supported by the Defense Science and Technology
Agency of Singapore under project agreement POD0613356.


\end{document}